\documentclass[pre,aps,superscriptaddress,twocolumn]{revtex4-1}
\usepackage{graphicx}
\usepackage[caption=false]{subfig}
\usepackage{epstopdf}
\usepackage{amsmath}
\usepackage{amssymb}
\usepackage[usenames,dvipsnames]{color}

\usepackage[normalem]{ulem}

\newcommand*{\kt}{k_\text{B}T}
\newcommand*{\nStar}{n_*}
\newcommand*{\Kpar}{\kappa_\parallel}
\newcommand*{\Kperp}{\kappa_\perp}
\newcommand*{\ebPar}{\epsilon^\text{b}_\parallel}
\newcommand*{\ebPerp}{\epsilon^\text{b}_\perp}
\newcommand*{\Es}{\epsilon_\text{s}}
\newcommand*{\Eb}{\epsilon_\text{b}}
\newcommand*{\Kb}{\kappa_\text{b}}
\newcommand*{\Np}{N_\text{p}}
\newcommand*{\Ebend}{E_\text{bend}}
\newcommand*{\muEq}{\mu_\text{eq}}
\newcommand*{\PhiStar}{\Phi_*}
\newcommand*{\va}{v_\text{a}}

\newcommand*{\nMax}{n_\text{max}}

\newcommand*{\Lt}{L}
\newcommand*{\LStar}{L_*}
\newcommand*{\KGZero}{K_\text{G}}
\newcommand*{\LFlat}{L_\text{flat}}

\begin{document}


\title{Thermodynamic size control in curvature-frustrated tubules: \\ Self-limitation with open boundaries}

\author{Botond Tyukodi}
\affiliation{Martin Fisher School of Physics, Brandeis University, Waltham, Massachusetts 02454, USA}
\author{Farzaneh Mohajerani}
\affiliation{Martin Fisher School of Physics, Brandeis University, Waltham, Massachusetts 02454, USA}
\author{Douglas M. Hall}
\affiliation{Department of Polymer Science and Engineering, University of Massachusetts, Amherst, Massachusetts 01003, USA}
\author{Gregory M. Grason}
\email{grason@mail.pse.umass.edu}
\affiliation{Department of Polymer Science and Engineering, University of Massachusetts, Amherst, Massachusetts 01003, USA}
\author{Michael F. Hagan}
\email{hagan@brandeis.edu}
\affiliation{Martin Fisher School of Physics, Brandeis University, Waltham, Massachusetts 02454, USA}

\pacs{--}

\begin{abstract}
We use computational modeling to investigate the assembly thermodynamics of a particle-based model for geometrically frustrated assembly, in which the local packing geometry of subunits is incompatible with  uniform, strain-free large-scale assembly.  The model considers discrete triangular subunits that drive assembly towards a closed, hexagonal-ordered tubule, but have geometries that locally favor negative Gaussian curvature. We use dynamical Monte Carlo simulations and enhanced sampling methods to compute the free energy landscape and corresponding self-assembly behavior as a function of experimentally accessible parameters that control assembly driving forces and the magnitude of frustration. The results determine the parameter range where finite-temperature self-limiting assembly occurs, in which the equilibrium assembly size distribution is sharply peaked around a well-defined finite size. The simulations also identify two mechanisms by which the system can escape frustration and assemble to unlimited size, and determine the particle-scale properties of subunits that suppress unbounded growth.
\end{abstract}

\date{\today}

\maketitle

The self-assembly of subunits into large, but finite-size, superstructures plays a central role in the functionalities, pathogenesis, and organization of biological systems (e.g. viral capsids
\cite{Zlotnick2011, Mateu2013, Bruinsma2015a, Perlmutter2015, Hagan2016, Twarock2018, Zandi2020, Hagan2021}, bacterial microcompartments \cite{Schmid2006, Iancu2007, Kerfeld2010, Rae2013, Chowdhury2014, Kerfeld2016} and other shelled cell organelles \cite{Sutter2008, Pfeifer2012, Nott2015, Zaslavsky2018}), and is becoming a route to design nanostructured assemblies for technological applications ~\cite{Sigl2021, Wagenbauer2017, Bale2016, Divine2021, Butterfield2017, King2014, Lai2014, Levasseur2021, Noble2016, Edwardson2018, Mosayebi2017}.  
In these examples self-limitation arises through `curvature control', meaning that subunits assemble with a preferred curvature that drives the structure to close upon itself, leaving no boundary for additional subunit association.

In a second class of self-limited, \emph{open-boundary} structures, self-assembly terminates at a well-defined equilibrium size without self-closure, leaving free boundaries at which subunits can readily exchange with the bulk. Possible examples with biological relevance include fiber bundles with well-defined diameters formed by sickle-cell hemoglobin or fibrin {\cite{Makowski1986, Weisel2004}}.  
Recently, it has been theoretically proposed that self-limited open-boundary assembly can be achieved through ‘geometric frustration’ (GF), in which the preferred local packing of subunits is incompatible with their preferred large-scale assembly structure \cite{Grason2016, Lenz2017, Meiri2021}. This incompatibility leads to a misfit strain energy that grows super-extensively with assembly size until it overwhelms the cohesive interactions that drive assembly, leading to a finite equilibrium (free energy minimum) size.  For example, theory showed that twisted fiber bundles can have self-limited diameters because the preferred skew of the filament-filament interactions is incompatible with the preferred 2D lattice packing in the cross-section {\cite{Hall2016, Weisel1987, Bruss2013, Haddad2019, Turner2003, Yang2010}}, while the growth of intra-membrane stretching with lateral size of 2D crystalline assemblies with incompatible Gaussian curvature can lead to ribbons of self-limited widths {\cite{Armon2014, Ghafouri2005, Schneider2005, Chen2016, Guo2014, Efrati2015, Roldan2021}}. In these and other examples, it is predicted that the range the self-limitation is delimited by mechanisms of `frustration escape', whereby defects or elastic distortions of the soft building blocks effectively screen the super-extensive cost of frustration and permit unlimited (bulk) assembly~\cite{Hall2017, Grason2016, Meiri2021}. 
 
The ability to understand and engineer the effects of GF on physical assemblies requires connecting the mechanisms and range of strain accumulations, as well as the complex modes of frustration escape, to particle-scale properties of the misfit building blocks.  However, to date our understanding of GF-limited assembly derives almost exclusively from continuum elastic descriptions of the ground-state energetics of aggregates within a limited set of a pre-assumed assembly morphologies.  It remains unclear when such models sample the thermodynamically relevant aggregates, whether and when finite-temperature fluctuations enhance or suppress self-limitation, and what is the physical limit to the range of assembly sizes and conditions where a given system exhibits self-limitation.    In this Letter, we study a discrete particle model of GF tubule assembly by combining dynamical assembly simulations with enhanced sampling to determine the equilibrium phase diagram of a GF system, accounting for finite temperature and concentration and without assumptions about assembly pathways, subunit packing geometry, or defects. We identify key experimentally accessible control parameters that enable tuning the finite equilibrium size, and we identify escape mechanisms, including defect formation and elastic shape-flattening. Further, we identify strategies to suppress frustration escape, thereby increasing the parameter range over which self-limited assembly can be achieved.

\begin{figure}
\begin{center}
\includegraphics[width=1.0\columnwidth]{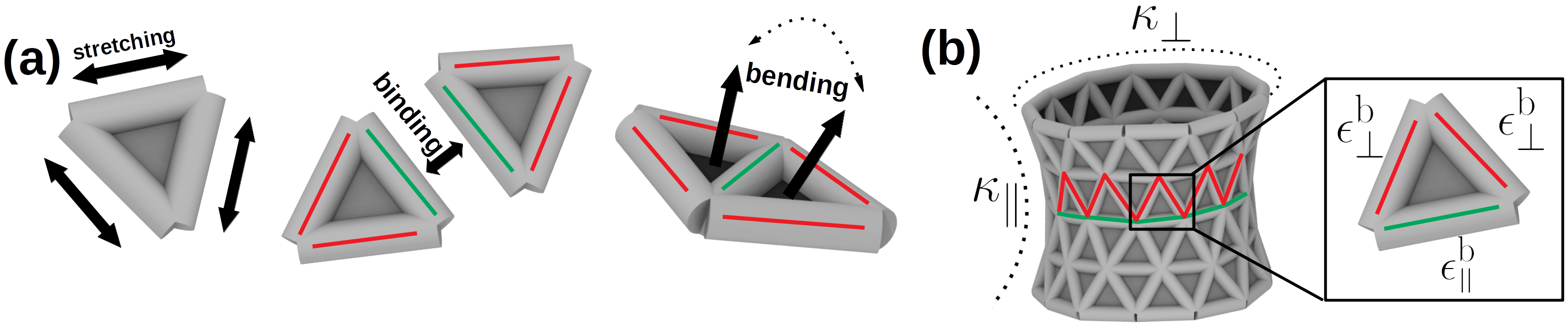}
\includegraphics[width=\columnwidth]{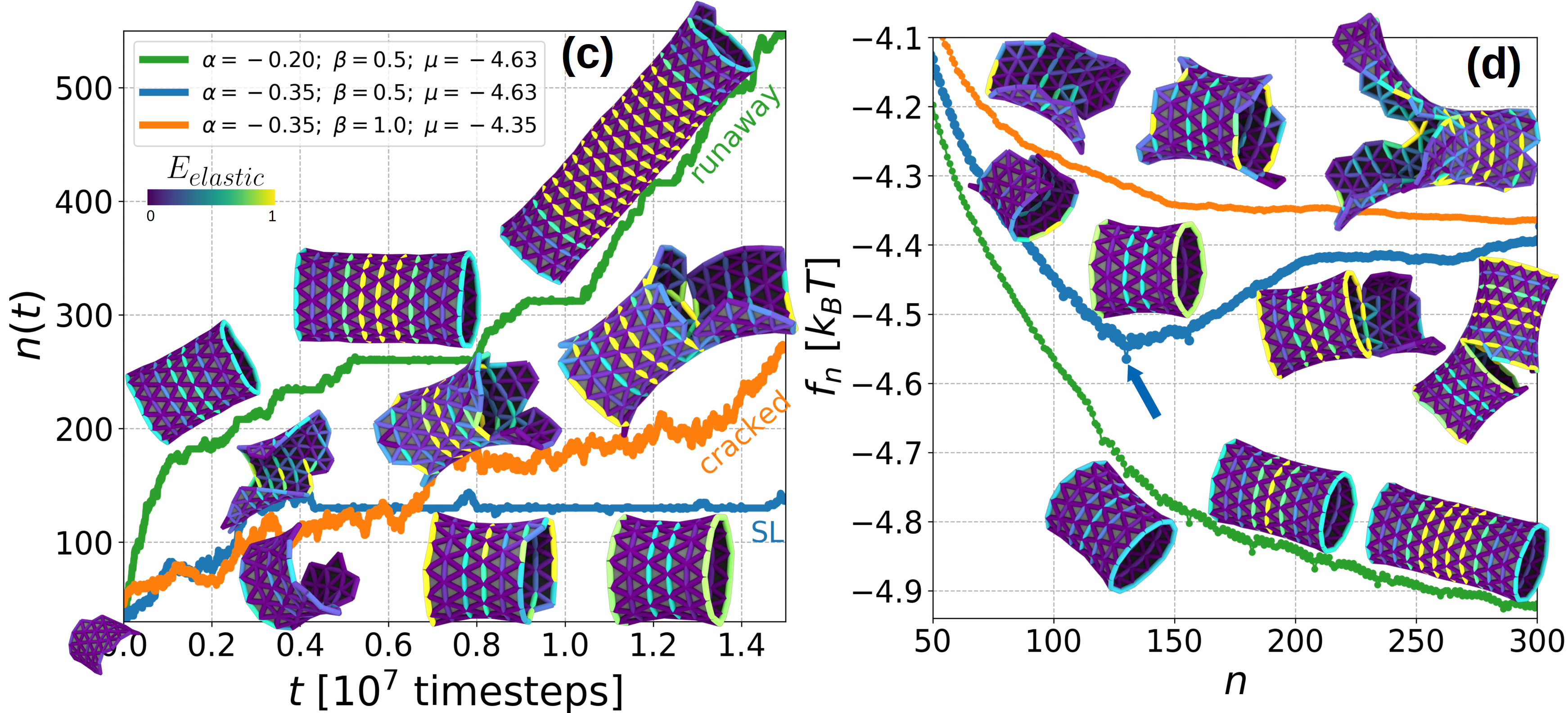}
\caption{
  \textbf{(a)} Model energies account for edge stretching, binding of subunits, and bending across bonds. \textbf{(b)} Subunits have three edge types and each type can only bind to its own kind. Two of the types (marked with red lines) have the same binding energy $\ebPerp$ and preferred dihedrals which set the concave curvature $\Kperp$. The third type (green lines) has a different binding energy $\ebPar$ and its preferred dihedral sets the convex curvature $\Kpar<0$. 
\textbf{(c)} Cluster size as a function of time for dynamical simulations that respectively illustrate the three classes of assembly behaviors: \textcolor{Orange}{cracked}, \textcolor{Green}{runaway} and \textcolor{NavyBlue}{self-limited} (with optimal size $\nStar=130$). Snapshots generated with Ovito \cite{Stukowski2010} along each trajectory are shown, with edges color-coded by their ground-state elastic energies. The curvature anisotropy $\alpha = \Kpar / \Kperp$, binding anisotropy $\beta = \ebPar / \ebPerp$ and chemical potential $\mu$ values are shown in the legend. \textbf{(d)} Per-subunit free energy profiles corresponding to the dynamical simulations  in (c), obtained by umbrella sampling with replica exchange between $36$ ensembles with different chemical potentials, temperatures, and bias potentials. The  regularly spaced local minima correspond to full ring closures. The arrow indicates the global minimum for the middle profile ($\alpha=-0.35, \beta=0.5$). The other profiles do not exhibit global minima.
For all 3 systems, the elastic parameters are $\Es= 150 \kt/l_0^2$ and $\Kb=300\kt$; the mean binding energy is  $\langle \Eb \rangle = (2 \ebPerp + \ebPar)/3 = -6.5\kt$; and the hoop curvature is $\Kperp \approx 2 \pi /(\Np l_0)$, which favors $2 \Np=26$ subunits in a ring. 
}
\label{fig:dynamical}
\end{center} 
\end{figure}

\textit{Model.} Inspired by recent DNA origami experiments \cite{Sigl2021} that target spherical capsids, we consider here a minimal model of GF assembly: a system of identical equilateral triangular subunits that assemble along their edges to form an elastic sheet curled into an axisymmetric tubular structure (Fig.~\ref{fig:dynamical}). The implications of GF for assembly studied here are generic, and analogous arguments can be made for systems with diverse sources of GF \cite{Grason2016,Lenz2017,Hagan2021}. In our model, the subunit edge shapes favor {\it concave} (i.e. self-closing) curvature $\Kperp>0$ in the hoop direction, but unlike cylindrical tubular assemblies \cite{Bollinger2019, Stevens2017} simultaneously favor {\it convex} curvature $\Kpar<0$ in the other.  Thus, this preferred dihedral geometry targets a preferred negative Gaussian curvature $\KGZero=\Kperp \Kpar<0$, which is incompatible with the uniform triangular lattice favored by the equilateral edge lengths of the subunits. The axisymmetric ground state configurations, determined by a balance of stretching and bending elasticity of the assembled sheet, correspond to catenoid-like tubules (`trumpets'). 

Frustration is driven by negatively-curved splaying at both free ends of a trumpet, which leads to a (hoop) strain that grows with trumpet length $\Lt$ as $\sim \KGZero \Lt^2$.  Zero-temperature continuum theory arguments, as considered previously~\cite{Ghafouri2005, Armon2014, Hagan2021}, present a simplified picture of the assembly thermodynamics.  For sufficiently short trumpets, the cumulative hoop stretching cost is much smaller than the unbending cost to flatten the axial curvature.  Hence, short trumpets adopt nearly the target Gaussian curvature, and the energy as a function of length is given by  the stretching and edge contributions, $F(L) \thicksim 2 \pi \kappa_\perp^{-1} \big(Y \KGZero^2 \Lt^5 + 2\Lambda\big)$ with $Y$ the 2D Young's modulus, and $\Lambda$ the line energy accounting for unsatisfied subunit interactions at the trumpet ends. An equilibrium self-limited length $\LStar$ in the canonical ensemble requires a minimum in the \emph{per-subunit} free energy $\thicksim F /( \Kperp^{-1} \Lt)$ \cite{Hagan2021}, which is given by $\LStar \cong \big(\Lambda/Y \KGZero^{-2} \big)^{1/5}$.  Unlike the aforementioned bundle and ribbon assemblies which undergo unfrustrated, unlimited growth in one direction, the simultaneously self-closing hoop and frustration-limited length of a trumpet implies a self-limitation with a thermodynamically finite mass, as we demonstrate below.

As edge tension increases, the optimal length grows, until the hoop stretching cost overwhelms the cost of unbending (flattening) in the axial direction, $\approx B \kappa_\parallel^2/2$, with $B$ the bending modulus.  This crossover from stretching- to bending-dominated costs defines a  characteristic length scale $\LFlat\thicksim \big(B / Y \Kperp^2\big)^{1/4}$ beyond which the trumpet interior `flattens' at the expense of bending energy due to deviations from the preferred curvature \cite{Hagan2021}. This simple picture suggests that for a range of elastic moduli and preferred curvatures, the system will achieve self-limited lengths dictated by the competition between stretching and line tension. However, there is a maximum achievable self-limited length $\LFlat$, beyond which the system escapes the superextensive penalties by flattening.


%
%
%
%

To investigate how the combined effects of finite temperature fluctuations as well as low-symmetry and defective morphologies alter this self-limiting assembly scenario, we adapt a discrete subunit model previously developed for icosahedral shell self-assembly  \cite{Rotskoff2018,Li2018b}. The model describes a growing elastic sheet comprised of identical triangular subunits, which each have three distinct edge types (Fig.~\ref{fig:dynamical}a). Here, we consider subunits that can bind to each other only along like edges, with binding energies $\Eb^{t}$ for edge-type $t=1,2,3$.  Two of the binding energies, corresponding to bonds within circumferential hoops, are set equal, $\Eb^{1} = \Eb^{2} \equiv \ebPerp$, while the third is different, $\Eb^{3} = \ebPar$, corresponding to axial bonding.  The in-plane stretching elasticity of the sheet derives from a Hookean energy penalty for deviations of each edge from its preferred length $l_0$ according to $E_\text{s} =\Es (l-l_0)^2/2$ with $\Es$ the edge's elastic modulus, and $l$ its instantaneous length. Here, we set $\Es$ the same for all three edge types. 
The preferred curvatures and bending modulus are controlled by a bending potential on dihedral angles along bonds according to $\Ebend = \Kb \big(\theta^{m} - \theta_0^{t(m)} \big)^2/2$, with  $m$ the index of a bound edge pair and $\theta_0^{t(m)}$ the preferred dihedral angle across bonds between bound edge pairs of type $t(m)$ (Fig.~\ref{fig:dynamical}b). Analogous to the binding energy, we set $\theta_0^{1} = \theta_0^{2} \equiv \theta_0^{\perp}$ and $\theta_0^{3} \equiv \theta_0^{\parallel}$, which respectively set the preferred principal curvatures $\Kperp$ and $\Kpar$.  In the limit of small preferred curvatures, the trumpet assemblies can be approximated as frustrated elastic shells, with bending and stretching elasticity and with target principal curvatures $\Kperp \approx \theta_0^\perp / \sqrt{3}$ and $\Kpar \approx {(2 \theta_0^\parallel + \theta_0^\perp)/\sqrt{3}}$. Details of the mapping are presented in the SI \cite{SIref}.

We performed dynamical simulations and free energy calculations using a grand canonical ($\mu VT$) Monte Carlo (MC) algorithm. To model the limit of dilute, noninteracting assemblies, we performed simulations of a single assembling structure in exchange with free subunits at fixed chemical potential $\mu$. The MC algorithm includes 11 moves (details are in the SI \cite{SIref, OpenMesh}) that account for subunit association/dissociation and structural relaxation of assembly intermediates. 
The model parameters are the stretching modulus $\Es$, the bending modulus $\Kb$, the binding energies $\ebPerp$ and $\ebPar$, the associated binding volume $\va$, the preferred curvatures $\Kperp$ and $\Kpar$, the chemical potential $\mu$ and the MC move attempt rates. Due to the large parameter space, here we set the subunit elastic properties to $\Es = 150 \kt / l_0^2$ and $\Kb = 300 \kt$, and the reference state and binding volumes to $v_0 = \va = l_0^3$. In addition, we fixed the average binding energy to $\langle \Eb \rangle = (2 \ebPerp + \ebPar)/3 = -6.5 \kt$ and the positive curvature  $\Kperp \approx 2 \pi / (\Np l_0 )$ so that $2 \Np=26$ subunits pack in an unstrained ring. In what follows, we consider the effects of varying two key dimensionless ratios: i) the curvature anisotropy: $\alpha = \Kpar/\Kperp$ and ii) binding anisotropy $\beta=\ebPar/\ebPerp$.


In addition to dynamical simulations, we investigated assembly thermodynamics by using umbrella sampling combined with parallel tempering to compute the equilibrium grand potential $\Omega_n$ as a function of cluster size $n$. The intra-cluster interaction free energy $F_n$ is then obtained by subtracting the ideal subunit translational free energy, $F_n = \Omega_n + \mu n$ \cite{Hagan2021, Frenkel1996, Earl2005, Ferguson2017, Yang2010, Kumar1992}. Technical details are in the SI \cite{SIref}.

\begin{figure}
\begin{center}
\includegraphics[width=0.95\columnwidth]{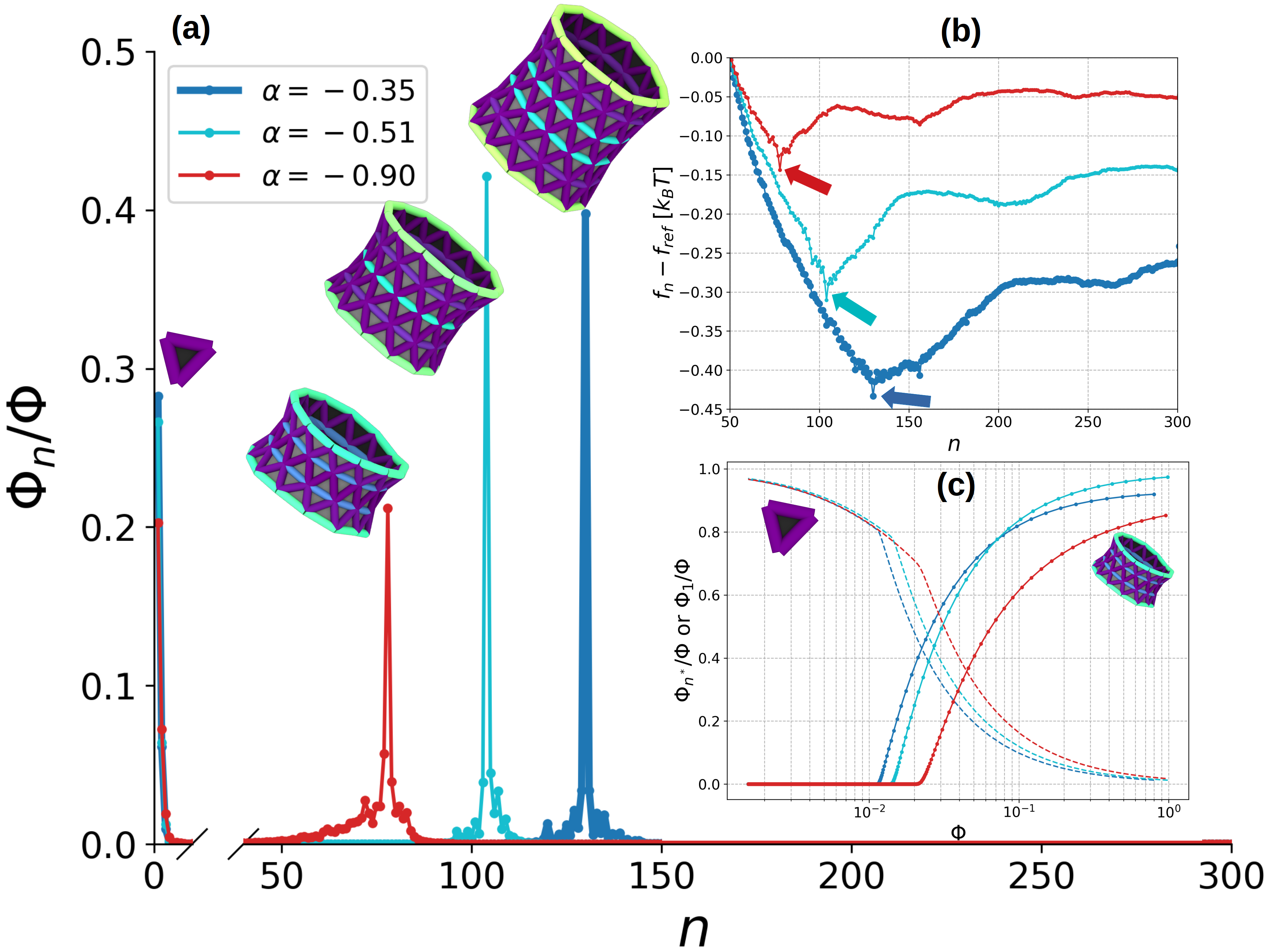}
\caption{Equilibrium cluster size distributions in the canonical ensemble ($NVT$) computed from free energy profiles.  \textbf{(a)} Distribution of subunits in $n$-clusters for $\beta=0.5$ and three $\alpha$ values. To aid visibility, the total concentration $\Phi$ for each curve is set to make the peaks at the self-limited size and free subunits both visible: $\alpha=-0.35$ ($\Phi=0.03$), $\alpha=-0.51$ ($\Phi=0.04$), $\alpha=-0.9$ ($\Phi=0.08$). \textbf{(b)} Per-subunit free energy profiles corresponding to the distributions in (a). Arrows indicate global minima ($\nStar$). 
\textbf{(c)} Fraction of subunits that are free monomers or in clusters close to the optimal size,  $n-\nStar \in [-10,10]$.   }
\label{fig:phi_n}
\end{center} 
\end{figure}

\textit{Results.}
Fig.~\ref{fig:dynamical}c shows example trajectories from dynamical simulations for three parameter sets, which illustrate the three structural categories that we observe \cite{SImovref}. At small curvature anisotropies (green line, labeled `runaway'), i.e. approaching an unfrustrated cylinder, $\alpha \rightarrow 0$, small structures form with the negative Gaussian curvature favored by individual subunits.  However, beyond a certain length the curvature-driven stresses are screened and the trumpet interior `flattens', pushing the negative Gaussian curvature to a narrow zone near the free ends. Consequently the length-dependent elastic energy of deformation approaches an extensive bulk term associated with interior flattening plus a length-independent boundary term, and the trumpet undergoes unbounded longitudinal growth.
At large curvature or binding anisotropies (orange line, `cracked'), the trumpets either fail to close or else crack, leading to disorganized (e.g. branched) but unbounded growth. However, we observe self-limited growth at an intermediate range of curvature anisotropy (blue line, `self-limited'). Structures have a catenoid-like geometry (negatively-curved regions of constant positive mean curvature unduloids), and after growing to final size $\nStar$ exhibit only small thermal fluctuations in length. 

To test whether dynamical simulations correspond to equilibrium phenomena, Fig. \ref{fig:dynamical}d shows the per-subunit interaction free energies $f_n$ for the three systems introduced in Fig. \ref{fig:dynamical}c.  The free energies of the runaway and cracked structures monotonically decrease with $n$, although they asymptotically flatten for large $n$ since the per-subunit edge energy diminishes as $1/n$. While both structures escape self-limitation, they do so via very different mechanisms. According to the continuum theory arguments above, for runaway trumpets the elastic energy is screened beyond a \textit{penetration length} $\LFlat\sim \Kperp^{-1/2}$, and flattening of the trumpet interior becomes energetically favorable. The elastic (bending) energy cost of flattening is extensive in $n$ and thus avoids the super-extensive cost of progressive stretching of outer hoops. In contrast, the cracks in high-curvature anisotropy structures incur line energy costs (from missing bonds), but locally release elastic energy. The elastic and cohesive energies of longitudinal cracks grow with their length, and hence, beyond the size at which cracks are stable, the trumpet energy grows extensively. Since either elastic flattening or crack formation preempt the self-limiting compromise between stretching and edge energy, they do not exhibit minima in $f_n$.

In contrast, the self-limited structure (blue line in Fig.\ref{fig:dynamical}d)  has a global minimum in $f_n$ at $\nStar=130$, which corresponds to a length of 5 rings. The additional local minima correspond to structures with 4, 6, and 7 closed rings respectively. Structures consisting entirely of closed rings are favored due to low edge energy. At larger sizes the free energy minimum structures exhibit cracks or correspond to multiple weakly-bound trumpets (each of which is self-limited). Note that structures corresponding to local minima in $f_n$ with $n<\nStar$ can be thermodynamically favored over the global minimum free energy structure at finite concentrations due to their greater translational entropy, whereas local minima with $n>\nStar$ are not favored at any concentration  (see Fig.~\ref{fig:phi_n} and Ref.~\cite{Hagan2021}).

From the free energies $f_n$, we compute the cluster size distribution (fraction of subunits in clusters of size $n$, $\Phi_n = n\exp[-n (f_n - \muEq) ]$) as a function of the total subunit density $\Phi$, where the equilibrium chemical potential $\muEq$ is determined by mass conservation $\Phi = \sum_n^{\nMax} \Phi_n$ \footnote{We set the maximum structure size to $\nMax=250$ because the umbrella sampling simulations sampled only up to $n=300$; the results are not sensitive to $\nMax$ provided that  $\nMax - \nStar$ is sufficiently large that $\Phi_{\nMax}\to 0$.}. Fig.~\ref{fig:phi_n} shows the resulting cluster size distributions and corresponding free energy profiles for three systems with different self-limiting sizes $n^*=78, 104, 130$, corresponding to a sequence of decreasing axial curvature anisotropies $\alpha$.  The increasing size of these minima are consistent with continuum theory expectations that the self-limiting trumpet lengths increase as axial curvature {\it decreases}, $L_* \sim \Kpar^{-2/5}$.  In each case the distribution is sharply peaked with a maximum at a size that is approximately equal to the per-particle free energy minimum size $\nStar$ and is insensitive to the total concentration $\Phi$. Analogous to classical aggregation, there is essentially no assembly below a pseudo-critical total subunit concentration $\PhiStar$; while for $\Phi \gg \PhiStar$ almost all subunits are in assemblies (with the monomer concentration approximately equal to $\PhiStar$). In contrast to classical aggregation but analogous to curvature-controlled self-limited assembly of spherical micelles or capsids \cite{Hagan2021}, the assembly size distribution remains narrowly distributed around the optimal size $\nStar$.

\begin{figure}
\begin{center}
\includegraphics[width=\columnwidth]{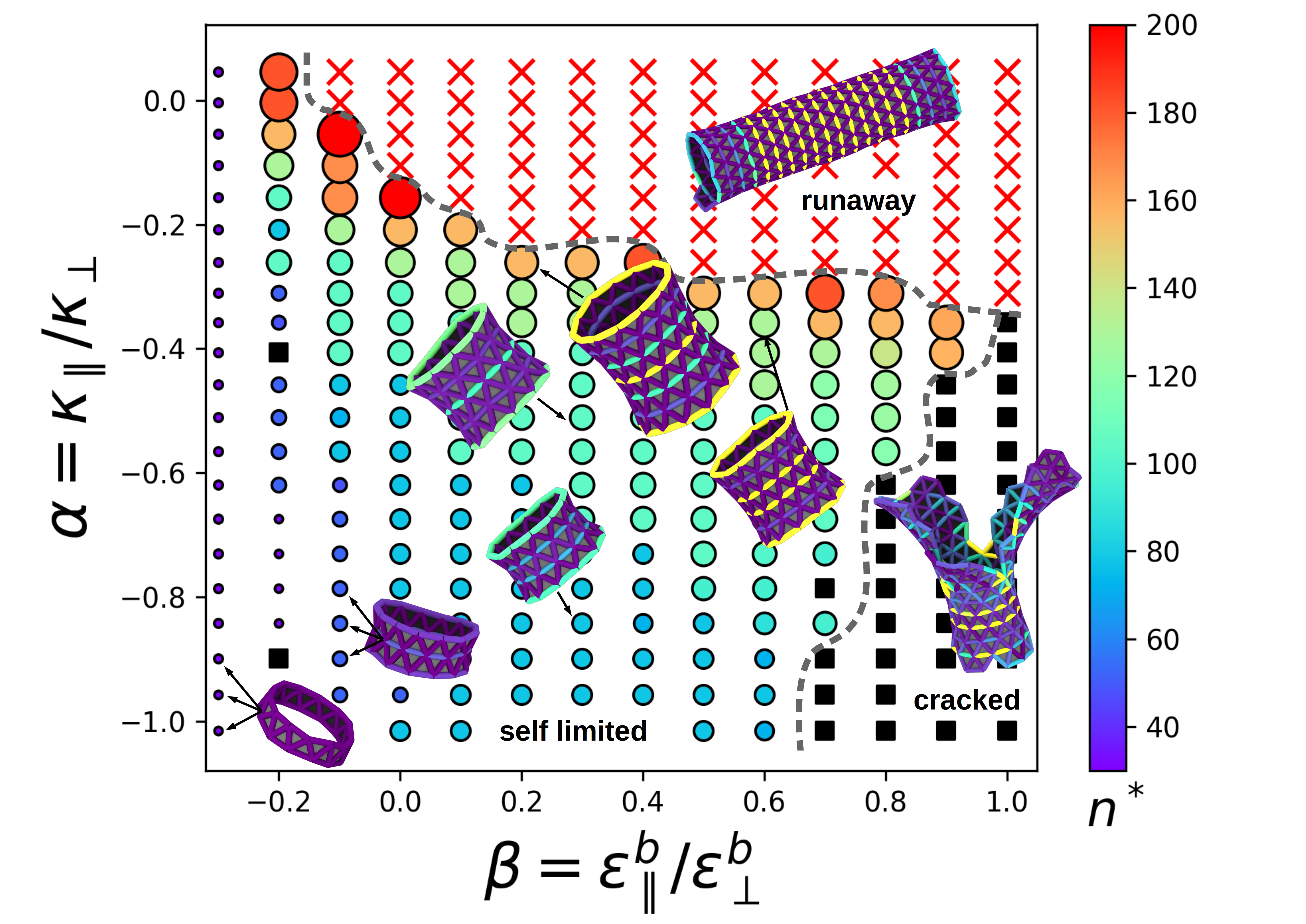}
\caption{Equilibrium phase diagram in the $\alpha-\beta$ plane computed from umbrella sampling. Symbols indicate phases: self-limited ($\circ$), cracked ($\blacksquare$), and runaway ($\times$). For the self-limited phase, symbols are colored and sized according to the optimal size $\nStar$.  Parameter sets are categorized as \textit{cracked} if  the coefficient of variation of the open boundary length is larger than $0.5$ at the lowest value of $f_n$; \textit{self-limited} if they are not cracked and exhibit a global minimum at $f_{\nStar}$ with $\nStar\le250$; and \textit{runaway} if they are neither cracked nor exhibit a global minimum for $n\le250$. Phase boundaries, which are drawn as guides to the eye, are qualitatively insensitive to these criteria.  The $\alpha=0$ and $\alpha=-1$ lines respectively correspond to preferred cylinder and minimal surface geometries, and  $\beta=1$ represents isotropic binding strength. 
}
\label{fig:phase}
\end{center} 
\end{figure}

Having established the existence of self-limiting equilibrium assembly at finite temperature and concentration, we now consider the parameters that determine this range and the self-limited size.  Fig.~\ref{fig:phase} shows the phase behavior and (in the self-limited regime) optimal size $\nStar$ computed from umbrella sampling, as a function of the dimensionless parameters controlling the frustration (negative curvature, $\alpha$) and binding anisotropy ($\beta$). We see that, notwithstanding the effects of thermal fluctuations and escape to low-symmetry morphologies, equilibrium self-limitation exists over a broad range of model parameters, with $\nStar$ increasing as $\alpha \to 0$, that is, as the negative curvature decreases and the preferred geometry approaches the unfrustrated cylinder. However, the curvature of the free energy profile near the minimum ($d^2 f_n/dn^2|_{\nStar}$) tends to decrease with $\nStar$ (see, for example, Fig.~\ref{fig:phi_n} inset), until the minimum disappears and the system enters the runaway phase. 

We also find that anisotropic binding, with lateral interactions stronger than inter-ring interactions ($\beta<1$), is essential for stable self-limited structures for all parameter ranges we have simulated. As $\beta \to 1$, the structure is sufficiently stabilized by strong inter-ring interactions that it becomes energetically favorable to break lateral interactions (i.e. form cracks) to relax the negative-curvature frustration.   Additionally, sufficiently strong inter-ring bonds overwhelm the flattening cost of axial unbending, and thus the threshold $\alpha$ value for runaway trumpets decreases with increasing $\beta$.  Hence, in a regime of $\beta \lesssim 1$, we observe that the self-limiting assembly is cut-off at the {\it upper} size range (low $|\alpha|$) by the elastic shape flattening mechanism, and at the {\it lower} size range (high $|\alpha|$) by the {\it inelastic} mechanism of longitudinal cracking.  As $\beta \to 1$, assembly transitions directly from cracked to runaway structures, without an intervening self-limiting regime.


\textit{Conclusions.}
In conclusion, we demonstrated that geometric frustration leads to equilibrium self-limitation over a finite, but specific, range of shape-misfit and binding anistropy, notwithstanding entropic effects at finite temperatures and complex, inelastic frustration escape mechanisms possible in a discrete particle model.  Importantly, the two dimensionless parameters that we focused on, targeted negative curvature and binding anisotropy, are experimentally controllable (e.g. with DNA origami \cite{Sigl2021, Bollinger2019, Wagenbauer2017} or polymer hydrogel particles {\cite{Bae2014, Kim2012}}). 
Finally, equilibrium self-limitation does not necessarily describe finite-time assembly; e.g. long-lived metastable intermediates (kinetic traps) may arise for far out-of-equilibrium initial states \cite{Hagan2014,Whitelam2015}. Thus,  more systematic dynamical simulations are needed to identify the self-limited regime under finite-time constraints.

\begin{acknowledgements}
 This work was supported by Award Number R01GM108021 from the National Institute Of General Medical Sciences (BT, FM, MFH), the Brandeis Center for Bioinspired Soft Materials, an NSF MRSEC, DMR-2011846 (BT, FM, DMH, GMH, MFH), and through NSF grant No. DMR-2028885 (DMH, GMG).
Computational resources were provided by NSF XSEDE computing resources (grant No. TG-MCB090163, Stampede, Comet, Expanse) and the Brandeis HPCC which is partially supported by DMR-2011486.
\end{acknowledgements}

\bibliographystyle{aipauth4-1}
\bibliography{trumpets_paper}
\bibliographystyle{unsrt}

\end{document}